\documentclass[amssymb,prl,twocolumn,floats,amsmath,showpacs,superscriptaddress]{revtex4}
\usepackage{bm}
\usepackage{graphicx}

\begin{document}


\title{Observation of strong-coupling effects in a diluted magnetic semiconductor Ga$_{1\textrm{-}x}$Fe$_{x}$N}
%
%
\author{W.~Pacuski}
\email{Wojciech.Pacuski@fuw.edu.pl}
\affiliation{Institute of Experimental Physics, Warsaw University,
Ho\.za 69, PL-00-681 Warszawa, Poland}

\affiliation{Institut N\'eel/CNRS-Universit\'e J. Fourier, Bo\^{\i}te Postale 166, F-38042 Grenoble Cedex 9, France}

\author{P.~Kossacki}
\affiliation{Institute of Experimental Physics, Warsaw University,
Ho\.za 69, PL-00-681 Warszawa, Poland}

\author{D.~Ferrand}
\affiliation{Institut N\'eel/CNRS-Universit\'e J. Fourier, Bo\^{\i}te Postale 166, F-38042 Grenoble Cedex 9, France}

\author{A.~Golnik}
\affiliation{Institute of Experimental Physics, Warsaw University,
Ho\.za 69, PL-00-681 Warszawa, Poland}

\author{J.~Cibert}
\affiliation{Institut N\'eel/CNRS-Universit\'e J. Fourier, Bo\^{\i}te Postale 166, F-38042 Grenoble Cedex 9, France}

\author{M.~Wegscheider}
\affiliation{Institute for Semiconductor and Solid State Physics, Johannes Kepler University, Altenbergerstrasse 69, A-4040, Linz, Austria}
\author{A.~Navarro-Quezada}
\affiliation{Institute for Semiconductor and Solid State Physics, Johannes Kepler University, Altenbergerstrasse 69, A-4040, Linz, Austria}
\author{A.~Bonanni}
\affiliation{Institute for Semiconductor and Solid State Physics, Johannes Kepler University, Altenbergerstrasse 69, A-4040, Linz, Austria}
\author{ M.~Kiecana}
\affiliation{Institute of Physics, Polish Academy of Sciences, al. Lotnik\'{o}w 32/46, PL 02-668 Warszawa, Poland}
\author{ M.~Sawicki}
\affiliation{Institute of Physics, Polish Academy of Sciences, al. Lotnik\'{o}w 32/46, PL 02-668 Warszawa, Poland}
\author{ T.~Dietl}
\affiliation{Institute of Physics, Polish Academy of Sciences, al. Lotnik\'{o}w 32/46, PL 02-668 Warszawa, Poland}
\affiliation{ERATO Semiconductor Spintronics Project of Japan Science and Technology Agency, al. Lotnik\'{o}w 32/46, PL 02-668 Warszawa, Poland}
\affiliation{Institute of Theoretical Physics, Warsaw University, ul. Ho\.za 69, PL-00-681 Warszawa, Poland}

%
\begin{abstract}
A direct observation of the giant Zeeman splitting of the free excitons in Ga$_{1\textrm{-}x}$Fe$_{x}$N is reported. The magnetooptical and magnetization data imply the ferromagnetic sign and a reduced magnitude of the effective $p$-$d$ exchange energy governing the interaction between Fe$^{3+}$ ions and holes in GaN, $N_0\beta^{\mathrm{(app)}} = +0.5\pm 0.2$~eV. This finding corroborates the recent suggestion that the strong $p$-$d$ hybridization specific to nitrides and oxides leads to significant renormalization of the valence band exchange splitting.
\end{abstract}


%
\maketitle


A strong hybridization between anion $p$ states and open
$d$ shells of transition metals (TM) is known to account for
spin-dependent properties of magnetic isolators, semiconductors, and
superconductors, such as antiferromagnetic superexchange,
hole-mediated Zener ferromagnetism, and the contribution of spin
fluctuations to Cooper pairing. Furthermore, the corresponding
exchange splitting of the bands gives rise to giant magnetooptical
and magnetotransport phenomena, a fingerprint of purposeful
spintronic materials. Studies of the $p$-$d$ exchange interaction
have been particularly rewarding in the case of Mn-based II-VI
dilute magnetic semiconductors (DMS) such as (Cd,Mn)Te
\cite{Furdyna:1988,Kacman:2001}. In these systems, Mn is an
isoelectronic impurity with a simple $d^5$ configuration, allowing a straightforward, quantitative determination of the $p$-$d$
exchange integral $\beta$ from the so-called giant Zeeman effect of
the exciton: using the virtual crystal (VCA) and molecular-field
approximations (MFA), one calculates a contribution to the spin
splittings proportional to the Mn magnetization and to $N_0|\beta|$, where $N_0$ is the cation density. At the same time, the
determination by photoemission spectroscopy \cite{Mizokawa:2002_a},
and the computation \cite{Zunger:1986,Larson:1988}, of band
structure parameters demonstrated that the antiferromagnetic $p$-$d$ exchange results indeed from $p$-$d$ hybridization.

According to the above insight, in II-VI oxides and III-V nitrides,
a small bond length and, thus, strong $p$-$d$ hybridization, should
result in particularly large values of $N_0|\beta|$, a prediction
supported by photoemission experiments \cite{Hwang:2005_a}.
Surprisingly, however, some of the present authors found abnormally
small exciton splittings in (Zn,Co)O \cite{Pacu06prb}, (Zn,Mn)O
\cite{Przezdziecka:2006_a}, and (Ga,Mn)N \cite{Pacuski:2007_a}.
Prompted by this contradiction, one of us suggested \cite{Diet07}
that due to the strong $p$-$d$ coupling, oxides and nitrides  form
an outstanding class of DMS, in which VCA breaks down, making the
\emph{apparent} exchange splitting $N_0\beta^{\mathrm{(app)}}$ small and of opposite sign. Importantly, the system bears some similarity
with semiconductor alloys such as Ga(As,N), so that experimental and theoretical studies of the valence band exchange splitting in the
strong coupling limit may significantly improve our understanding of these important alloys.

To check the above model, we have carried out high-resolution
studies of magnetization and magnetoreflectivity in the free exciton region for (Ga,Fe)N epitaxial layers, thoroughly characterized
previously \cite{Bona07}. Since in GaN, in contrast to ZnO, the actual ordering of valence subbands is settled, the sign of $N_0\beta^{\mathrm{(app)}}$ can be unambiguously determined from polarization-resolved magnetooptical spectra. Furthermore, unlike Mn, Fe in GaN is an isoelectronic impurity with the simple $d^5$ configuration \cite{Bona07,Baur94Malguth06}, allowing a straightforward interpretation of the data. Our results lead to a value of  $N_0\beta^{\mathrm{(app)}} = +0.5 \pm 0.2$~eV, which provides an important experimental support for the theory \cite{Diet07}.

\begin{figure}[bt]
\includegraphics[width=\linewidth]{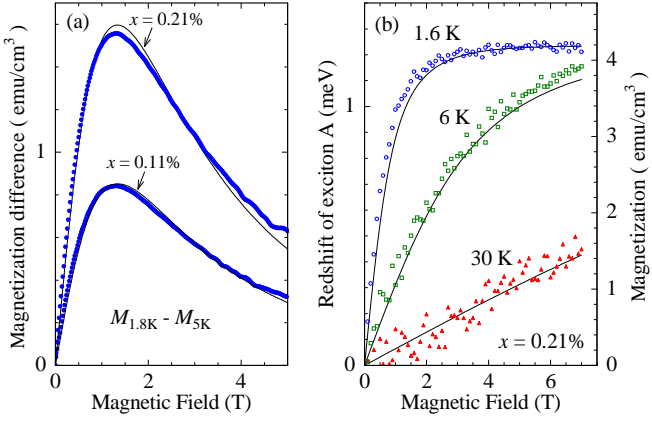}
\centering \caption[]{(color online) (a) Difference between magnetizations
measured (symbols) at 1.8 and 5~K. Same difference, computed (solid
line) using the Brillouin function for $S = 5/2$ and treating the Fe concentration $x$ as the only fitting parameter. (b) Comparison
between the computed magnetization (solid lines, right axis) and the redshift of exciton \emph{A} in $\sigma^-$ polarization (symbols,
left axis) for the $0.21$\% sample at three temperatures. The
spectra are shown in Fig.~\ref{fig:reflectivity}(a).}
\label{fig:brillouin}
\end{figure}

The 0.7~$\mu$m thick layers of Ga$_{1\textrm{-}x}$Fe$_{x}$N were grown \cite{Bona07}
by metalorganic vapor phase epitaxy (MOVPE), on [0001] sapphire
substrates with a 1~$\mu$m thick, wide-band gap (Ga,Al)N buffer
layer, which is transparent in the free exciton region of
Ga$_{1\textrm{-}x}$Fe$_{x}$N. The Fe flow rate was adjusted to keep
the Fe content well below $0.4$\%, which is the solubility limit of
Fe in GaN under our growth conditions. According to detailed
luminescence, electron paramagnetic resonance, and magnetic
susceptibility studies \cite{Bona07}, in this range the Fe ions
assume mostly the expected Fe$^{3+}$ charge state corresponding to
the $d^5$ configuration, for which the spin polarization as a
function of temperature $T$ and magnetic field $B$ is determined by
the Brillouin function B$_S(T,B)$ with spin $S = 5/2$ and Land\'e
factor $g_{Fe^{3+}} = 2. 0$. This is confirmed by our magnetooptical data shown in
Fig.~\ref{fig:brillouin}, which scale with B$_{5/2}(T,B)$. For the reported samples, we determine Fe content $0.11 \pm 0.02$\% and $0.21\pm 0.02$\% by fitting the difference in magnetization values measured at 1.8 and 5.0~K up to 5~T in a high-field SQUID magnetometer, Fig.~\ref{fig:brillouin}(a).

%

\begin{figure}[bt]
\includegraphics[width=\linewidth,clip]{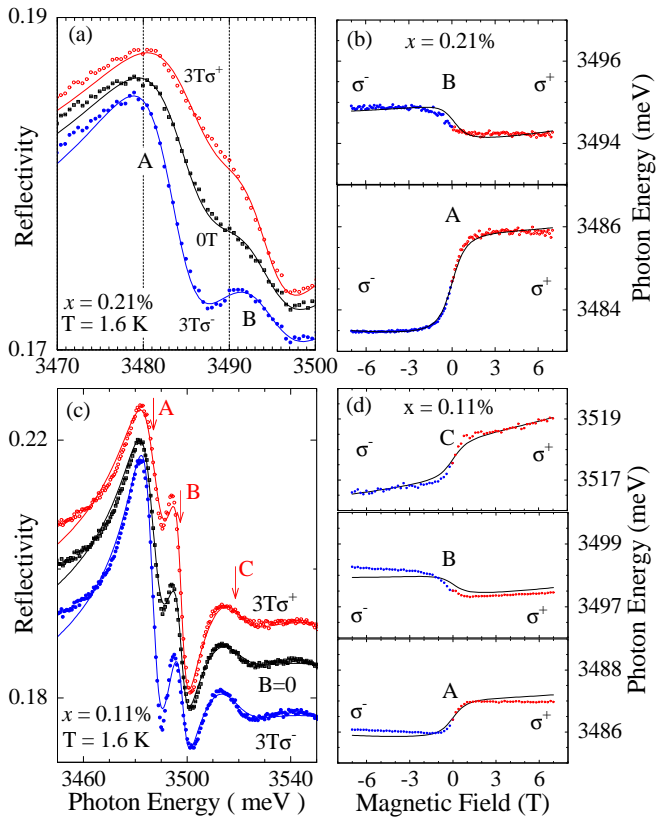}
\centering \caption[]{(color online) Reflectivity of Ga$_{1\textrm{-}x}$Fe$_{x}$N
in Faraday configuration at 1.6~K  for $x=0.21$\% (a) and 0.11\% (c)
where particularly well resolved excitons $A$, $B$, and $C$ are visible in
$\sigma^-$ polarization. (b,d) Field-induced exciton shifts of the
excitons determined from the polariton model (points) compared to
the expectations of the exciton model (solid lines). The determined
values of the exciton and exchange parameters are shown in Table I.
} \label{fig:reflectivity}
\end{figure}

Magneto-reflectivity spectra were collected in the Faraday
configuration (propagation of light and magnetic field along the
normal to the sample, which is the $c$-axis of the wurtzite
structure), and the light helicity $\sigma^\pm$ defined with respect to the magnetic field direction. Owing to the high quality of the layers, either two, or the three, free excitons of GaN \cite{Step99}, $A$, $B$, $C$, are resolved, and their Zeeman shifts are visible in the spectra (Fig.~\ref{fig:reflectivity}). In $\sigma^-$
polarization, the $A-B$ splitting increases and the $B-C$ splitting
decreases with the field. Opposite shifts are observed in $\sigma^+$.
Since these shifts are entirely different from those observed in
pure GaN \cite{Step99}, and related to the magnetization
[Fig.~\ref{fig:brillouin}(b)], we conclude that the Fe ions create a "giant Zeeman effect" in (Ga,Fe)N. Remarkably, however, the shift
observed for $A$ exciton is opposite in sign and significantly smaller at given $x$ than in other wurtzite DMS with $d^5$ ions such as (Cd,Mn)Se \cite{Arciszewska:1986}.


We now describe the methodology that we employed to
extract the values of the apparent $p$-$d$ exchange energy
$N_0\beta^{\mathrm{(app)}}$ for the valence band and the apparent
$s$-$d$ exchange energy $N_0\alpha^{\mathrm{(app)}}$ for the
conduction band, from the reflectivity spectra. It involves two
major steps. First, we calculate the reflection coefficients as a
function of the photon energy, using a polariton model which
incorporates the ground states of excitons $A$ and $B$
\cite{Pacu06prb}, but also exciton $C$, excited states of excitons,
and the continuum. Second, the field dependence of the exciton
energies is calculated taking into account the essential features
of the exciton physics in wide-band gap wurtzite DMS
\cite{Pacu06prb,Pacuski:2007_a}.


The starting point of the polariton model is the
dielectric function $\epsilon(\omega,k)$ containing polariton poles
corresponding to excitons $A$ and $B$ \cite{Pacu06prb,Lago77}: by
solving analytically Eq.~3 of Ref.~\onlinecite{Lago77}, we obtain
the refractive index and then the reflection coefficient to be
compared to experimental spectra. However, we replace the background dielectric function $\epsilon^*_0$ by the residual dielectric
function $\epsilon^*(\omega)$, which takes into account additional
contributions which are expected to exhibit no significant polariton effect \cite{Step97}: exciton $C$, excited states of $A$, $B$ and
$C$ \cite{Step97} which are optically active ($S$-states),
and transitions to the continuum of unbound states \cite{Tang95}.
Hence
\begin{eqnarray}
\epsilon^*(\omega) = \epsilon^*_0+
4\pi\alpha_{0C}\frac{\omega_C}{\omega_{C}^2-\omega^2-i\omega\Gamma_{C}}+
\nonumber
\\
+\sum_{j=A,B,C}\left( \sum_{n=2}^{\infty}
\frac{4\pi\alpha_{0j}}{n^3}\frac{\omega_{n,j}}{\omega_{n,j}^2-\omega^2-i\omega\Gamma_{n,j}}+\epsilon_{j,\mathrm{ub}}\right).
\end{eqnarray}
Here $\epsilon^*_0 =5.2$ \cite{Step97}; $\alpha_{0j}$, $\omega_j =
E_j/\hbar$, and $\Gamma_j$ are the polarizability, resonant
frequency, and damping rate of each exciton $A$, $B$, and $C$,
treated as adjustable parameters (found to be close to those reported in Ref.~\onlinecite{Step97}). Resonant energies of the excited
states $n$ are $\hbar\omega_{n,j}=E_j+R^*_j-R^*_j/n^2$, where
$R^*_j$ are the effective Rydbergs known from our and other
\cite{Korn99} studies of excitons in GaN. The corresponding damping
parameters are calculated by using an empirical formula
\cite{Toul80,Korn99},
$\Gamma_{n,j}=\Gamma_{\infty}-(\Gamma_{\infty}-\Gamma_{j})/n^2$,
with one common damping rate $\Gamma_{\infty}$, which is an
additional fitting parameter. Finally,  the contribution
$\epsilon_{j,\mathrm{ub}}$ from unbound states is given in Eq.~5 of
Ref.~\onlinecite{Tang95}, with the amplitude determined by the
exciton polarizabilities $\alpha_{0j}$ and the damping parameter
$\Gamma_{\infty}/2$ \cite{Serm98,Step97}. Calculated reflectivity
spectra are compared to experimental ones in Figs.~2(a,c), and the
exciton energies are displayed as a function of the magnetic field
in Fig.~2(b,d).

In the second step, the field dependance of the exciton energies is
calculated for the hamiltonian
\begin{equation}
 H=E_0 + H_{\mathrm{v}} + H_{\mathrm{e-h}} +H_{\mathrm{Z}}+H_{\mathrm{diam}}+H_{sp-d}^{\mathrm{(app)}},
 \end{equation}
where $E_0$ is the band-gap energy and $H_{\mathrm{v}}$ describes the top of the valence band in semiconductors with the wurtzite
structure \cite{Bir74,Gil95,Juli98}. This term includes the trigonal
component of both crystal field and biaxial strain (described by the parameter $\tilde\Delta_1$) as well as the anisotropic spin-orbit
interaction (characterized by $\Delta_2$ and $\Delta_3 = 5.5$~meV \cite{Gil95} for the direction parallel and perpendicular to the $c$-axis, respectively). The component $H_{\mathrm{e-h}}$ describes the electron-hole interaction within the exciton \cite{Juli98,Step99,Pacuski:2007_a,Pacu06prb} and it involves the effective Rydberg and the electron-hole exchange integral $\gamma = 0.6$~meV \cite{Juli98}. Effects linear in the magnetic field are taken into account by the standard Zeeman hamiltonian $H_{\mathrm{Z}}$ \cite{Step99,Bir74}, parameterized by the effective Land\'e factor $g_e=1.95$ for the electrons and the relevant Luttinger effective parameter $\tilde\kappa=-0.36$ that describes the splitting of all three valence subbands \cite{Step99}. The diamagnetic shift is described by a single term quadratic in the magnetic field, $H_{\mathrm{diam}}= dB^2$, where ${d=1.8~\mu}$eV/T$^2$ \cite{Step99}. Finally, $H_{sp-d}^{\mathrm{(app)}}$ is the exchange interaction between Fe ions and carriers in the extended states from which the excitons are formed. We use the standard form of the $s$,$p$-$d$ hamiltonian \cite{Furdyna:1988,Kacman:2001,Arciszewska:1986}, which is proportional to the scalar product of the carrier spin and the
magnetization. However, being aware that these splittings can have a different meaning from the usual one \cite{Diet07,Sliw07}, we use
effective quantities $\alpha^{\mathrm{(app)}}$ and
$\beta^{\mathrm{(app)}}$. The  field dependence of the exciton
energies, calculated with the values of fitting parameters
collected in Table I, is compared with the experimental results in
Fig.~2(b,d).

\begin{table}

\begin{tabular}{cccccc}
$x$ & $N_0(\beta^{\mathrm{(app)}}$$-$$\alpha^{\mathrm{(app)}})$& $N_0(\beta^{\mathrm{(app)}}$+$\alpha^{\mathrm{(app)}})$& $E_A$ &$\tilde\Delta_1$ & $\Delta_2$  \\
\hline
$0.11\%$& $0.44\pm0.2$ & $0.60\pm0.2$ & 3.4864 &$19.4$& $7.5$   \\
$0.21\%$& $0.46\pm0.1$ & $0.63\pm0.3$& 3.4846 &$15.5$& $7.2$   \\
\end{tabular}
\caption{\label{tab}
Fitting parameters describing the exciton energies and the giant Zeeman effect shown in Fig.~\ref{fig:reflectivity}(b,d). Units are eV, except for $\tilde\Delta_1$ and $\Delta_2$, which are given in meV.}
\end{table}


The detailed analysis of this hamiltonian \cite{Pacuski:2007_a}
shows that the giant Zeeman shifts of excitons $A$ and $B$ are
mainly governed by
$N_0\beta^\mathrm{(app)}$$-$$N_0\alpha^\mathrm{(app)}$. Moreover,
exciton $A$, which in GaN is formed from valence band states with
parallel spin and orbit ($\Gamma_9$ state), shifts to low energies
in $\sigma^-$ polarization: hence the exchange integral difference
$N_0\beta^\mathrm{(app)}$$-$$N_0\alpha^\mathrm{(app)}$ is positive.
At the same time, exciton $B$ is mixed with exciton $C$, whose shift is primarily controlled by
$N_0\beta^\mathrm{(app)}$+$N_0\alpha^\mathrm{(app)}$. Hence, by
incorporating exciton $C$ in our description of the reflectivity
spectra, we can evaluate independently the apparent exchange energy
$N_0\beta^{\mathrm{(app)}}$ for the valence band and the apparent
exchange energy $N_0\alpha^{\mathrm{(app)}}$ for the conduction
band. This determination is quite accurate in the case of the 0.11\%
sample, for which exciton $C$ is spectrally well resolved, but an
estimation is still possible in the 0.21\% sample through its effect
on exciton $B$. From the shifts of $B$ and $C$, we find that the
sign of $N_0\beta^\mathrm{(app)}$+$N_0\alpha^\mathrm{(app)}$ is also
positive, and its value is quite close to that of
$N_0\beta^\mathrm{(app)}$$-$$N_0\alpha^\mathrm{(app)}$, see Table I.
Hence, $|N_0\alpha^\mathrm{(app)}|$ is much smaller than
$|N_0\beta^\mathrm{(app)}|$, and $N_0\beta^\mathrm{(app)}$ is
positive. More precisely, the complete fit yields the values of the
exchange energies $N_0\beta^{\mathrm{(app)}} = +0.5 \pm 0.2$~eV and
$N_0\alpha^{\mathrm{(app)}} =+0.1 \pm 0.2$~eV.


We note that our evaluation of ${N_0\alpha^{\mathrm{(app)}}}$
includes within the experimental error the values of $N_0\alpha
=0.25 \pm 0.06$~eV found in early studies of Mn-based II-VI DMS
\cite{Kacman:2001}. We have no reason to question here the
applicability of the standard description of the conduction band in (Ga,Fe)N --
contrary to (Ga,Mn)As and (Ga,Mn)N, for which the magnitudes of
$N_0\alpha^{\mathrm{(app)}}$ were found to be reduced under some experimental conditions \cite{Sliw07}.


For the expected $d$ level arrangement, both the ferromagnetic sign
and the small magnitude of the apparent $p$-$d$ exchange energy are
surprising. Indeed, in GaN, the Fe $d^5/d^6$ acceptor-like level
resides less than 3~eV above the top of the valence band
\cite{Baur94Malguth06}. This state, $e_g$$\downarrow$, and also the
higher lying $t_2$$\downarrow$ level that can hybridize with the
valence band states, remain unoccupied in intrinsic (Ga,Fe)N. At the same time, no donor-like $d^5/d^4$ state has been found within the
GaN gap. This could be expected, as in the TM series a particularly
large correlation energy $U$ separates the $d^5$ and $d^6$ shells.
Hence, the occupied Fe $t_2$$\uparrow$ levels reside within the
valence band. According to the Schrieffer-Wolf theory, in such a
case the $p$-$d$ exchange coupling is antiferromagnetic: this was
confirmed by magnetooptical studies of tellurides and selenides
containing either Mn or Fe, including studies carried out in our
labs, which lead systematically to $N_0\beta = -1.4 \pm 0.5$~eV
\cite{Kacman:2001}.

However, it has been recently remarked \cite{Diet07} that for an
appropriately strong TM potential, like the one expected for oxides
and nitrides, the TM ion can bind a hole -- a trend which was already
suggested by strong deviations from the VCA in (Cd,Mn)S \cite{Benoit1992} and by the analysis of {\em ab-initio} calculations \cite{Bouzerar}. A summation of infinite series of relevant
self-energy diagrams demonstrates that in such a situation, the spin
splitting of extended states involved in the optical transitions
remains proportional to magnetization of the localized spins, but
the apparent exchange energy becomes significantly renormalized \cite{Diet07}. In
fact, for the expected coupling strength, the theory predicts $ -1 <
\beta^{\mathrm{(app)}}/\beta < 0$, as observed here for (Ga,Fe)N.


A fruitful comparison can be made with the modification of the
conduction band of GaAs induced by a slight doping with Nitrogen
\cite{Shan99Klar00}. When a hydrostatic pressure is applied, the
Nitrogen isoelectronic centers create localized states, which are
observed in photoluminescence, but also extended states (the
so-called $E^+$ states) to which the oscillator strength is
transferred so that they are observed in reflectivity; moreover,
these optically active states exhibit an anticrossing with the
localized states, and the strength of the anticrossing increases
with the Nitrogen density. As a result, the transition to these
$E^+$ states exhibits a shift to \emph{high energy} when the density of \emph{low energy states} increases. In a DMS with a large value
of $N_0|\beta|$, only the TM impurities with the right spin orientation (antiparallel to the hole spin) are expected to form a localizing
center \cite{Benoit1992,Diet07}: hence the giant Zeeman shift in (Ga,Fe)N can be understood as resulting from a similar anticrossing but in this case the density of relevant localized centers is additionally controlled by
the field-induced orientation of the localized spins.


In conclusion, giant Zeeman splitting has been observed by
magnetoreflectivity for the $A$, $B$, and $C$ excitons in
Ga$_{1-x}$Fe$_{x}$N. The spectra are well described by the exciton
model valid for DMS with the wurtzite structure.  The determined
sign and magnitude of the \emph{apparent} $p$-$d$ exchange energy
${N_0\beta^{\mathrm{(app)}}=+0.5\pm0.2}$~eV constitutes an important
verification of a recent theory \cite{Diet07}, which describes the
effects of the $p$-$d$ interaction  circumventing the
virtual-crystal and molecular-field approximations that break down
in nitrides and oxides. In these systems, TM ions bind holes,
precluding in this way the occurrence of carrier-mediated
ferromagnetism in $p$-type materials. However, at sufficiently high
hole densities, an insulator-to-metal transition is expected. In the
metallic phase, many-body screening of local potentials annihilates
bound states. Large spin-splitting and robust ferromagnetism are
expected in this regime \cite{Dietl:2002,Popescu:2007}.


This work was supported by Polish Ministry of Science and Higher Education (project N202 006 31/0153), by the Austrian Fonds zur {F\"{o}rderung} der wissenschaftlichen Forschung - FWF (projects P17169-N08 and N107-NAN)
and by the French Ministry of Foreign Affairs.

%

\end{document}